\documentclass[prb,twocolumn,showpacs]{revtex4}
\usepackage{graphicx}

\begin{document}
\title{Detecting percolative metal-insulator transition in manganites by resistive relaxation }
\author{X. J. Chen$^{1,2}$, H.-U. Habermeier$^{2}$, and C. C. Almasan$^{1}$ }
\affiliation{$^{1}$Department of Physics, Kent State University, Kent, Ohio 44242\\
$^{2}$Max-Planck-Institut f\"{u}r Festk\"{o}rperforschung, D-70569 Stuttgart, Germany }
\date{\today}

\begin{abstract}
We report an experimental study of the time dependence of resistivity of a 
La$_{0.9}$Sr$_{0.1}$MnO$_{3}$ ultrathin film in order to elucidate the underlying mechanism 
for metal-insulator transition and colossal magnetoresistance CMR effect. There is a clear 
change of sign in the resistive relaxation rate across the metal-insulator transition driven 
by temperature or magnetic field. When measuring in increasing temperature or decreasing 
magnetic field, the resistivity increases with time in the metallic state but 
decreases with time in the insulating state. These relaxation processes 
indicate that the metal-insulator transition and the associated CMR are a direct result of 
phase separation and of percolation of the metallic phase.
\end{abstract}

\maketitle

Mixed-valence manganites of R$_{1-x}$A$_{x}$MnO$_{3}$ (R is a rare-earth trivalent element, 
and A a divalent dopant) are a particularly important class of materials because of their 
scientific interest and potential technological applications.\cite{coey} For an 
intermediate doping region, an insulator-metal transition occurs at a temperature $T_{IM}$, 
marked by a peak in the electrical resistivity $\rho$. The system is insulating 
($d\rho/dT<0$) above $T_{IM}$ and metallic ($d\rho/dT>0$) below $T_{IM}$. This phase change 
is accompanied by a transition from high-temperature paramagnetic (PM) to low-temperature 
ferromagnetic (FM) state at the Curie temperature $T_{C}$. Under the application of an 
external magnetic field, $T_{IM}$ shifts to a higher temperature and the resistivity is 
strongly suppressed. This gives rise to a huge magnetoresistance, called colossal 
magnetoresistance (CMR).

Although the theoretical understanding of the CMR phenomenon is still incomplete, 
double-exchange,\cite{zene} electron-phonon coupling,\cite{mill} and orbital effects
\cite{maez} are commonly recognized as its main ingredients. Contrasting ground states
can result from the combination of all these internal interactions. A subtle energy
balance between some of the competing states may lead to the formation of electronic 
phase mixtures. Recent studies \cite{gork,vare,mayr,heff,savo,dete,ueha,fath,zhan}
have provided accumulating evidence for the existence of phase separation in manganites. 
For example, it has been proposed theoretically \cite{gork,mayr} that percolation should 
play an essential role in the transport properties. Scanning tunneling spectroscopy 
\cite{fath} and magnetic force microscopy \cite{zhan} have revealed the formation and 
evolution of percolation networks of metallic and insulating phases across the 
metal-insulator transition.

In this paper, we use resistive relaxation measurements to reveal the nature of the
metal-insulator transition and CMR effect of La$_{0.9}$Sr$_{0.1}$MnO$_{3}$, a compound which 
belongs to a composition range where the occurrence of phase separation has recently been 
proposed theoretically.\cite{okam} We discover resistive relaxation in both the insulating 
and metallic regions, with opposite relaxation rates in these two regions and a change of 
sign across the metal-insulator transition, driven by temperature and magnetic field. These 
experimental data imply phase separation in both insulating and metallic regions. 
They also indicate that the metal-insulator transition and the associated colossal 
magnetoresistance effect are the result of percolation of the metallic clusters imbedded 
into the insulating matrix. 

The 50 $\AA$ La$_{0.9}$Sr$_{0.1}$MnO$_{3}$ films used in our experiments were grown on a 
(100) SrTiO$_{3}$ substrate with a pulsed laser deposition technique described previously.
\cite{chen} Because the lattice constant of SrTiO$_{3}$ is smaller than that of
La$_{0.9}$Sr$_{0.1}$MnO$_{3}$, the film grown on SrTiO$_{3}$ is subject to compressive
strain, resulting in a suppression of the charge and orbital ordering (CO) transition 
temperature $T_{CO}$ as well as a significant enhancement of $T_{IM}$ compared to the bulk.
\cite{urus,para} The epitaxially strained growth of these ultrathin films has been revealed 
by high-resolution transmission electron microscopy. The detailed structure analysis is 
presented elsewhere.\cite{chen1} Resistivity and magnetoresistivity 
measurements in zero field and under various magnetic fields up to 14 T have been performed 
through a conventional four probe method by using a Quantum Design Physical Properties 
Measurement System. The temperature stability was better than $0.01\% $ during the relaxation 
measurements.

The temperature dependence of the resistivity $\rho(T)$ measured in different applied magnetic 
fields is shown in Fig. 1. All these data are characterized by hysteresis, which is 
exemplified in the figure by the zero-field resistivity curves. Upon cooling from 400 K,
the $\rho(T)$ curve exhibits a peak at $T_{IM}=300$ K, signaling the insulator-metal transition.
At a lower temperature $T_{CO}\approx 50$ K, $\rho(T)$ displays a minimum, indicating the
appearance of a low-temperature FM insulating-like state for $T<T_{CO}$. In this low
temperature state, new structural reflections characteristic of CO appear in neutron
\cite{yama} and x-ray \cite{endo} diffraction patterns. In warming from 2 K, the $\rho(T)$ 
curve is first below the cooling one, it overshoots the cooling curve around 305 K and
exhibits a peak (metal-insulator transition) at $T_{MI}=308$ K. Such a behavior gives a 
clear indication of intrinsic phase inhomogeneity \cite{ueha,zhan}. An applied magnetic
field $H$ suppresses the resistivity, shifts $T_{MI}(H)$ towards higher temperatures 
(see inset to Fig. 1), and, hence, gives rise to the CMR effect.

\begin{figure}[tbp]
\begin{center}
\includegraphics[width=\columnwidth]{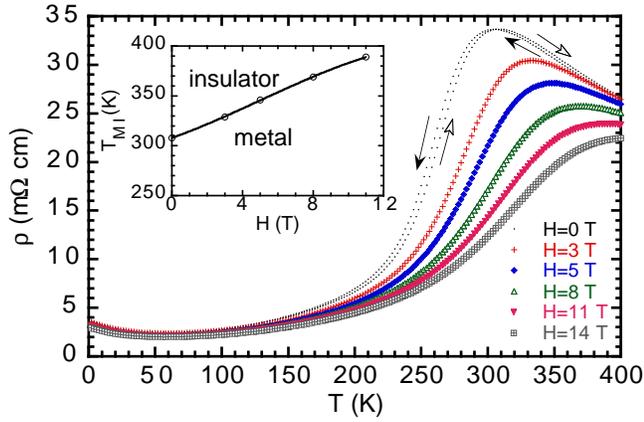}
\end{center}
\caption{Temperature dependence of the resistivity for a 50 $\AA$ 
La$_{0.9}$Sr$_{0.1}$MnO$_{3}$ ultrathin film under zero-field and various magnetic fields measured
in a warming run. The zero-field data show a typical resistivity hysteresis, with the arrows
indicating the directions of the temperature sweeps. Inset: Temperature -- magnetic field phase 
diagram. The solid line is the metal to insulator transition boundary. }
\end{figure}

We measured the magnetization as a function of temperature up to 300 K in a magnetic field of
0.5 T using a Quantum Design Superconducting Quantum Interference Device magnetometer. The 
magnetization is still large ($\sim$70 emu/cm$^{3}$) at 300 K compared to its saturation value
of $\sim$300 emu/cm$^{3}$ at low temperatures. This indicates that the Curie temperature $T_{C}$ 
of this film is above 300 K. A lower $T_{IM}$ value compared to $T_{C}$ has been previously
reported in La$_{1-x}$Sr$_{x}$MnO$_{3}$ thin films.\cite{chen,gonz} A possible explanation for 
$T_{IM}<T_C$ is the existence of microscopic phase segregation, with FM clusters embedded in 
PM insulating matrix. The FM clusters are large enough to give a magnetic contribution, but do 
not percolate for $T>T_{IM}$. Our relaxation data shown below support such a scenario.

Relaxation studies of the resistivity have proven to be an useful tool to investigate the  
dynamics of the competing superexchange and double-exchange interactions in manganites. 
\cite{vonh} The relaxation effect on the resistivity and magnetization of the half-doped 
manganites R$_{0.5}$A$_{0.5}$MnO$_{3}$ has been studied recently in order to address
nonequilibrium phenomena present in these systems.\cite{kawa,smol,ueha1,lope,levy} Here, 
we use relaxation studies of the resistivity to elucidate the mechanism responsible for 
the metal-insulator transition in manganites. Figure 2 shows representative profiles of the 
relaxation of the resistivity measured in the presence of a magnetic field $H=3$ T at different 
temperatures in a warming run. The film was initially cooled to 2 K in the presence of the 
magnetic field. The temperature was then increased to the desired value at which the 
resistive relaxation measurement was performed. Then, the temperature was stabilized to the 
next desired value and the resistive relaxation measurement performed again. The resistivity 
increases with time for $50 \leq T \leq 320$ K and decreases with time 
for $T\geq 330$ K. Figure 3 shows the relaxation of the resistivity measured at a fixed 
temperature $T=320$ K for various magnetic fields in a decreasing field run. The resistivity 
increases with time for $H\geq 3$ T and decreases with time for $H\leq 2$ T. As shown in the 
inset to Fig. 1, the temperature of $320$ K is the metal-insulator transition temperature 
for a field of about $2.2$ T. Thus the system is in the metallic state for $H\geq 3$ T and 
$T\leq 320$ K, and in the insulating state for $H \leq 2$ T and $T \geq 320$ K. The 
decrease (increase) of the resistivity with time in the insulating (metallic) state
is, therefore, obtained through both protocols: measuring the resistive relaxation at different 
$T$ while keeping $H$ fixed or at different $H$ while keeping $T$ fixed.

\begin{figure}[tbp]
\begin{center}
\includegraphics[width=\columnwidth]{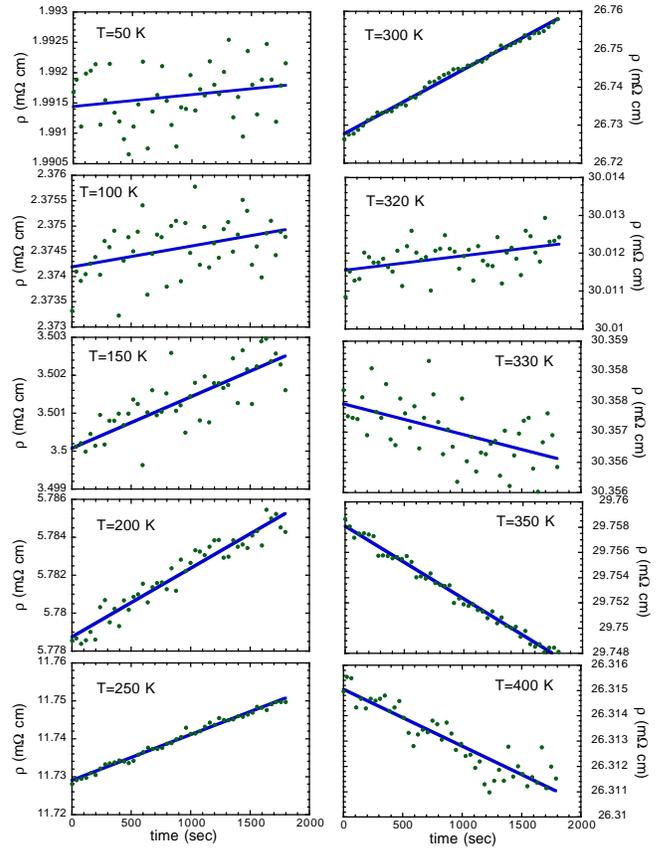}
\end{center}
\caption{Time dependence of the resistivity for a 50 $\AA$ 
La$_{0.9}$Sr$_{0.1}$MnO$_{3}$ ultrathin film measured in a magnetic field of $H=3$ T
at different temperatures, as labeled on the graph, in a warming run. The solid lines are 
linear fits of the data.  }
\end{figure}

We performed a linear fit of all relaxation data with $\rho(t)=\rho(t_{0})+\eta t$. Here $t$ 
is the relaxation time, $\rho(t_{0})$ is the initial resistivity, and $\eta$ is the resistive 
relaxation rate, which is the only free parameter. Figures 4(a) and 4(b) show the variation 
of $\eta$ with $T$ measured in $H=3$ T, and with $H$ measured at $T=320$ K, respectively. As 
shown above, $\eta$ is positive in the metallic state, zero around the metal-insulator
transition, as marked by the solid line and dashed region, and negative in the insulating
state. The resistive relaxation rate has a maximum in the metallic state around 280 K,
as shown in Fig. 4(a). The inset to this figure shows that the absolute value of CMR 
for the same field of 3 T is maximum around the same temperature. Also, both $\eta$ (see
Fig. 4(b)) and the CMR (see Fig. 1), measured at a constant temperature, increase with 
increasing $H$ in the metallic state. These results indicate that there is a direct
correlation between the behavior of the colossal magnetoresistance and the relaxation 
rate of the resistivity.

\begin{figure}[tbp]
\begin{center}
\includegraphics[width=\columnwidth]{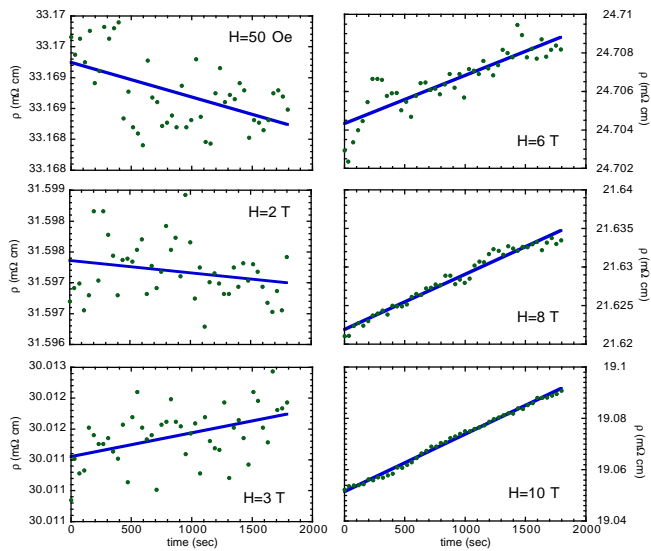}
\end{center}
\caption{ Time dependence of the resistivity of a 50 $\AA$ La$_{0.9}$Sr$_{0.1}$MnO$_{3}$ 
ultrathin film measured at a temperature of 320 K for different applied magnetic fields, as 
labeled on the graph, in a field decreasing run. The solid lines are linear fits of the data. }
\end{figure}

It is also interesting to note that $\eta$ decreases with decreasing temperature for $T<280$ 
K and approaches zero again at $T=50$ K (see Fig. 4(a)). Recalling that there is a minimum 
in the $\rho(T)$ curve at $T_{CO}\approx 50$ K (see Fig. 1), the relaxation of the 
resistivity also reveals the transition from the FM metallic state to the FM insulating-like 
state below 50 K.

One may consider whether the value of $\eta$ includes a contribution which is a result of 
temperature drift. To verify this point, we calculated the values of $\partial \rho/\partial t$ 
(hence $\eta$) which would be the result of temperature drift, from $\partial \rho/\partial T $ 
determined from $\rho$ vs $T$ curves in a magnetic field of 3 T and $\partial T/\partial t $ 
determined from the time dependence of $T$ recorded during the resistive relaxation measurements. 
These values of $\eta$ are two orders of magnitude smaller than those determined from the 
resistive relaxation measurements shown in Fig. 4(a). Thus, the contribution to $\eta$ from 
temperature drift is negligible.

\begin{figure}[tbp]
\begin{center}
\includegraphics[width=\columnwidth]{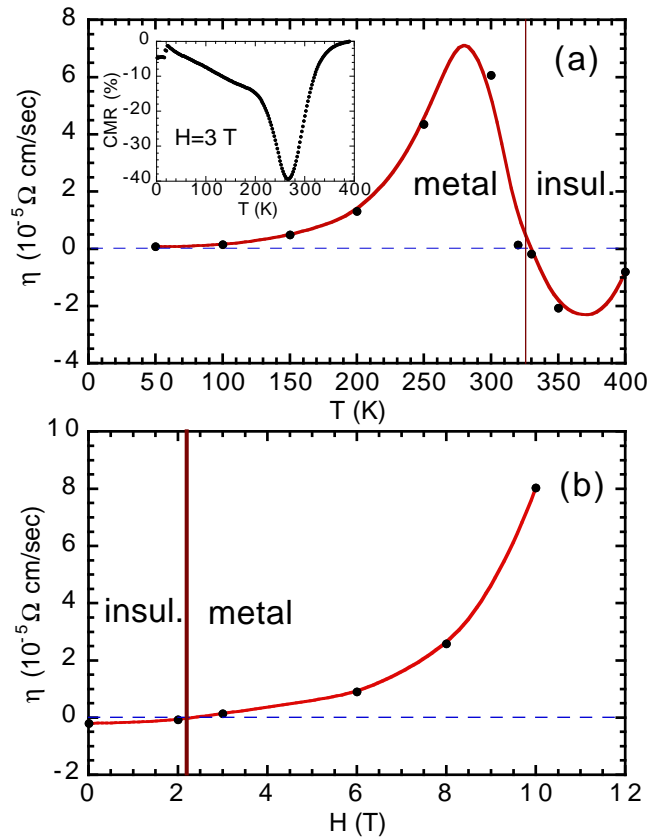}
\end{center}
\caption{ (a) Temperature dependence of the relaxation rate of the resistivity $\eta$ obtained 
from resistivity vs time measurements in a magnetic field of 3 T for a 50 $\AA$
La$_{0.9}$Sr$_{0.1}$MnO$_{3}$ ultrathin film. The vertical line marks the metal-insulator 
transition temperature for a magnetic field of 3 T. Inset: Temperature dependence of the 
magnitude of the colossal magnetoresistance CMR $=(\rho(3T)-\rho(0))/\rho(0)$ for a magnetic 
field of 3 T. (b) Relaxation rate of the resistivity $\eta$ as a function of magnetic field 
H measured at a temperature of 320 K. The dashed region marks the metal-insulator transition 
corresponding to this temperature. The dashed lines are guides to the eye. }
\end{figure}

Next we demonstrate that these resistive relaxation measurements point towards the scenario
of phase separation and percolation in this compound. Specifically, these data show that 
the metal-insulator transition and the CMR effect of manganites are a direct result of 
intrinsic phase inhomogeneity over the whole measured temperature range and of the evolution 
of the relative volume of the insulating and metallic phases with temperature or magnetic 
field. The presence of resistive relaxation up to 400 K indicates that there are FM clusters 
in the PM region. The existence of FM clusters above $T_{C}$ has been detected by small-angle 
neutron scattering measurements.\cite{dete} These conductive FM clusters are embedded into an
insulating host matrix. Upon lowering temperature or increasing magnetic field, the FM 
clusters grow and the ratio of FM metal to PM insulator becomes larger and larger. At a 
critical temperature or magnetic field, the conducting FM domains become interconnected
across the whole sample, the percolation is established. Therefore, the resistivity exhibits
a large decrease and the insulator-metal transition takes place. By further lowering 
temperature or increasing applied magnetic field, the fraction of the FM insulating phase 
decreases rapidly; accordingly, the fraction of the FM metallic phase increases. The 
existence of these electronically phase separated FM insulating and FM metallic phases 
in the low-temperature FM metallic region has been confirmed by measurements of M\"{o}ssbauer 
spectroscopy,\cite{chec} muon spin relaxation,\cite{heff} and nuclear magnetic resonance.
\cite{savo}

When the system is warmed from low temperatures or field decreased from high values, the 
fraction of the FM metallic domains does not decrease as fast as it increased on the cooling 
run.\cite{zhan} Therefore, this fraction is larger than its equilibrium value for a certain
temperature or magnetic field on both sides of $T_{MI}$. This gives rise to the hysteretic 
behavior of the resistivity both below and above  $T_{MI}$ and to a larger value of the
metal-insulator transition temperature, as depicted in Fig. 1. Also, since the fraction of 
the FM metallic domains is larger than its corresponding equilibrium value, with increasing time 
the resistivity in the metallic region increases, the metal-insulator transition temperature 
(the temperature corresponding to the threshold for percolation) shifts to lower temperatures 
(since the threshold for percolation does not change in time), and the resistivity in the 
insulating region decreases. As a result, the resistive relaxation rate is positive in the  
metallic state and negative in the insulating state with a change in sign across the 
metal-insulator transition, as shown by Fig. 4.

We should emphasize that the results presented are representative for the CMR manganites which 
have metal-insulator transitions, and not only for ultrathin films of La$_{0.9}$Sr$_{0.1}$MnO$_{3}$. 
We have found similar resistive relaxations in both insulating and metallic regions of thicker 
films of the same composition and of bilayer La$_{1.2}$Sr$_{1.8}$Mn$_{2}$O$_{7}$ single 
crystal manganites. 

In summary, our experimental data of La$_{0.9}$Sr$_{0.1}$MnO$_{3}$ ultrathin films clearly 
show opposite resistive relaxation rates in the paramagnetic insulating state and the 
ferromagnetic metallic state, with a crossover in the sign of the relaxation rate around the 
metal-insulator transition. The experimental results provide strong support for the existence 
of phase separation in both insulating and metallic regions. Our findings, of a remarkable 
consistency between the effects of temperature and magnetic field, suggest that the 
metal-insulator transition and the negative colossal magnetoresistance are a result of the
percolation of metallic ferromagnetic domains.

We thank G. Cristiani for growing the thin films, H. Zhang and C. L. Zhang for help in
transport measurements, and M. Varela and J. Santamaria for TEM studies of our films. The
work at KSU was supported by NSF Grant No. DMR-0102415.

\end{document}